\documentclass[11pt,twoside]{article}
\usepackage{newpasp}

\usepackage{epsf}

\index{Zabludoff, A.I.}

\begin{document}

\title{Gas and Galaxy Evolution in Poor Groups of Galaxies}
\author{Ann I. Zabludoff}
\affil{University of Arizona and Steward Observatory}


\begin{abstract}
Poor groups of galaxies are the repositories of most of the baryons
in the local Universe and are environments in which galaxy evolution
is likely to be both more recent and simpler than in the cores of rich
clusters.  Yet we know little about groups outside of our own Local
Group.  We discuss recent results from optical, X-ray, and radio
observations of nearby poor groups that focus on the evolution of gas
and galaxies in these common environments.


\end{abstract}




\section{Introduction}

Poor groups are both typical environments for galaxies and relatively
simple environments for galaxy evolution, especially in contrast to the complex
mechanisms that might drive galaxy evolution in hotter, denser
clusters of galaxies.  We define poor groups as systems with five or
fewer L$^*$ or brighter galaxies.  Many galaxies, including our own
Milky Way, lie in poor groups, which can be roughly divided into three
major classes based on their X-ray and galaxy morphologies.  The first
class includes the NGC 533 group (Figure 1), which
is marked by a large fraction of early type galaxies and by luminous,
symmetric, diffuse X-ray emission coincident with the central giant
elliptical.  Groups of the second class are like Hickson Compact
Group 90 (Figure 1), which has a lower fraction of early types
and an apparently unrelaxed X-ray halo not clearly associated
with particular galaxies.  In HCG 90, we observe several interacting
galaxies in the core instead of a giant elliptical.  The third class
of poor groups includes our own Local Group, which consists of two
giant late type galaxies and their satellite populations apparently
falling together for the first time.  These groups have
few or no early type members, have little or no
diffuse X-ray emission, and, while bound, are not necessarily
virialized.

A classic problem in the study of poor groups has been that, with five
or fewer known members, it is difficult to differentiate true, bound systems
from chance superpositions of galaxies along the line-of-sight.
Furthermore, even in bound groups, the small number of known members
has prevented an accurate calculation of the group kinematics.  We
use multi-object spectroscopy to determine whether there are members
further down the luminosity function and find a surprising result:
groups of the first class, those with X-ray luminous halos, can have
$50-60$ members down to $M^* + 5$ projected within $\sim 0.5 h^{-1}$ Mpc, their
virial radii (Figure 2).  These newly
discovered galaxies confirm such groups as bound systems and provide a
means for assessing the group's dynamical state.  Groups of the second and
third classes, which have at most marginally detected X-ray emission,
are not as easy to identify as real.  The populations of dwarf
galaxies present in the X-ray luminous groups are not observed in
X-ray faint groups (Figure 2).  Deeper spectroscopy is required to
establish whether these poorer groups have fainter dwarfs, analogs to
those galaxies in the Local Group fainter than the Small Magellanic
Cloud.

\begin{figure}
\plottwo{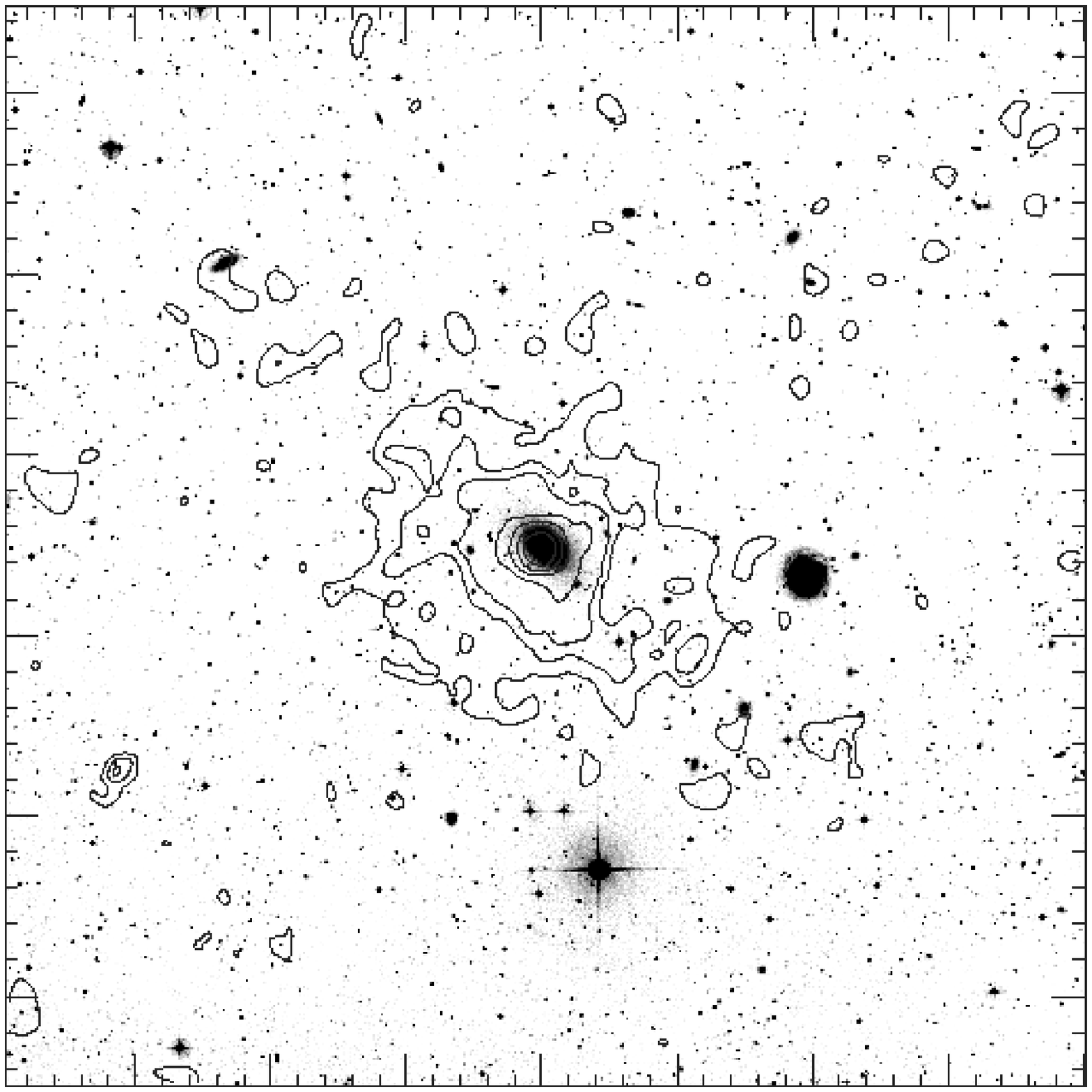}{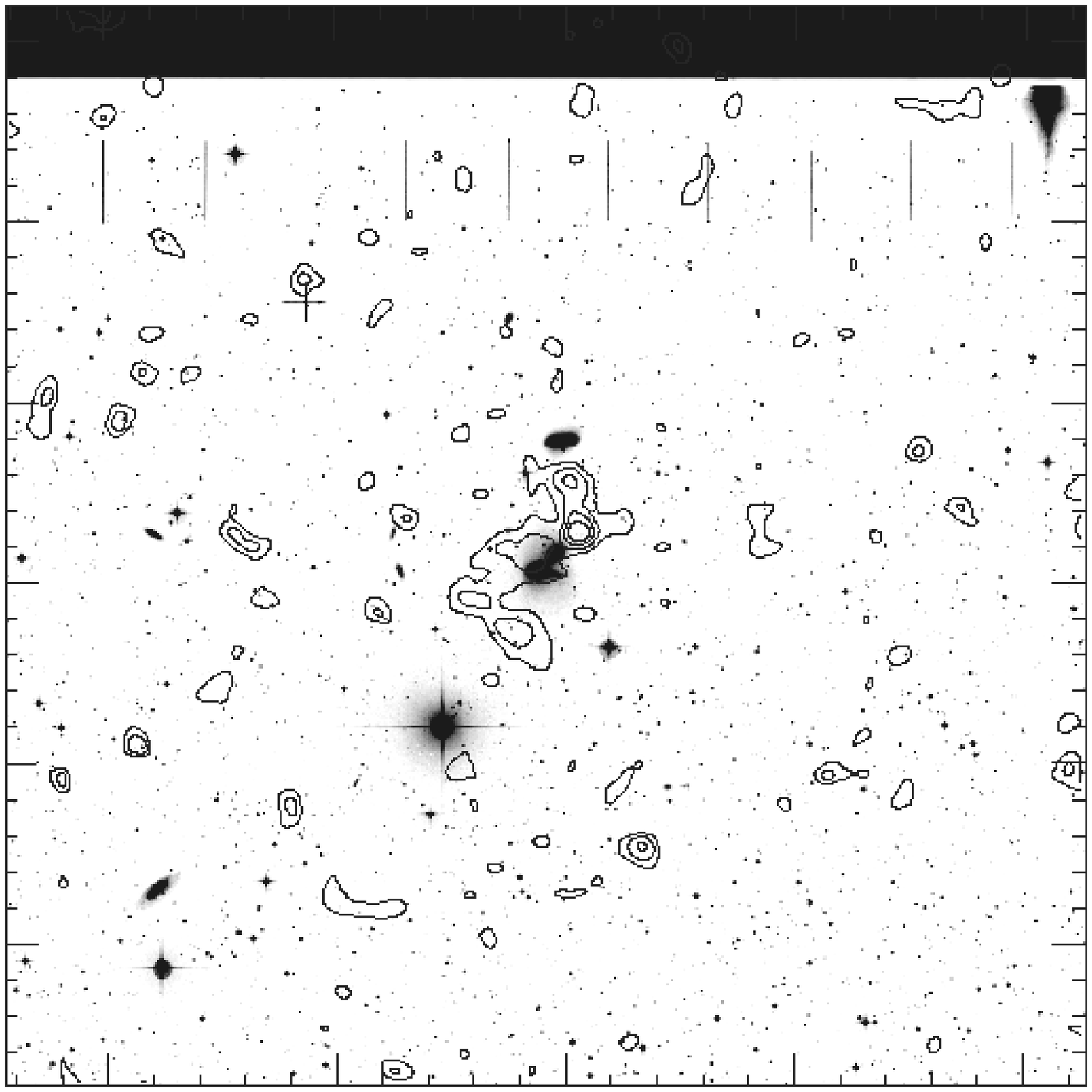}
\caption{\small {\it Left:}  A Digital Sky Survey image
of the poor group NGC 533 (greyscale) with the ROSAT X-ray emission (contours) superimposed.  
Note the symmetric X-ray emission, the central,
giant elliptical, and the dearth of late type galaxies.  
{\it Right:}  A corresponding image of the poor group Hickson 
90.  In this group, the X-ray emission is not smooth, symmetric, or even apparently associated with
the galaxies.  Instead of one dominant central elliptical, 
there are several interacting galaxies, including late types, in the core.
}
\label{fig1}
\end{figure}

\begin{figure}
\plotone{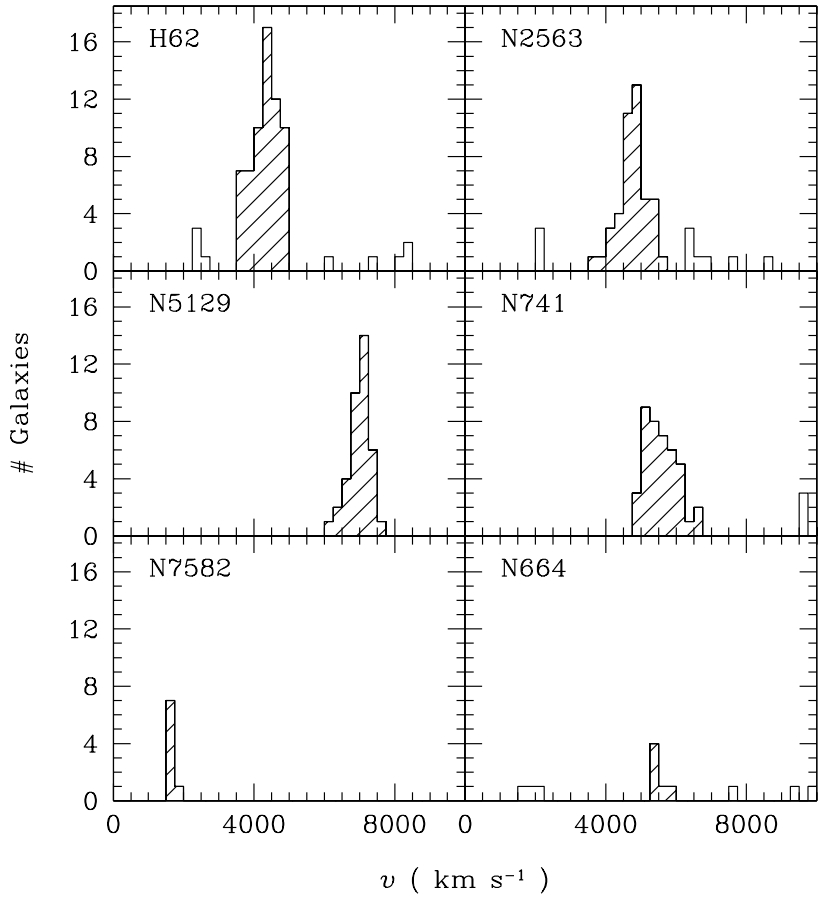}
\caption{\small Galaxy velocity histograms for six poor groups.
The first four have diffuse, luminous
X-ray halos, the last two are not detected in X-rays.  The non-detected groups also have fewer
members, and thus we are unable to confirm that they are real, bound systems instead of chance superpositions
of galaxies along the line-of-sight.  We note, however, that the properties of
the Local Group, which is bound (but not virialized), are consistent with those of the non-detected groups.
}
\label{fig2}
\end{figure}

\begin{figure}
\plottwo{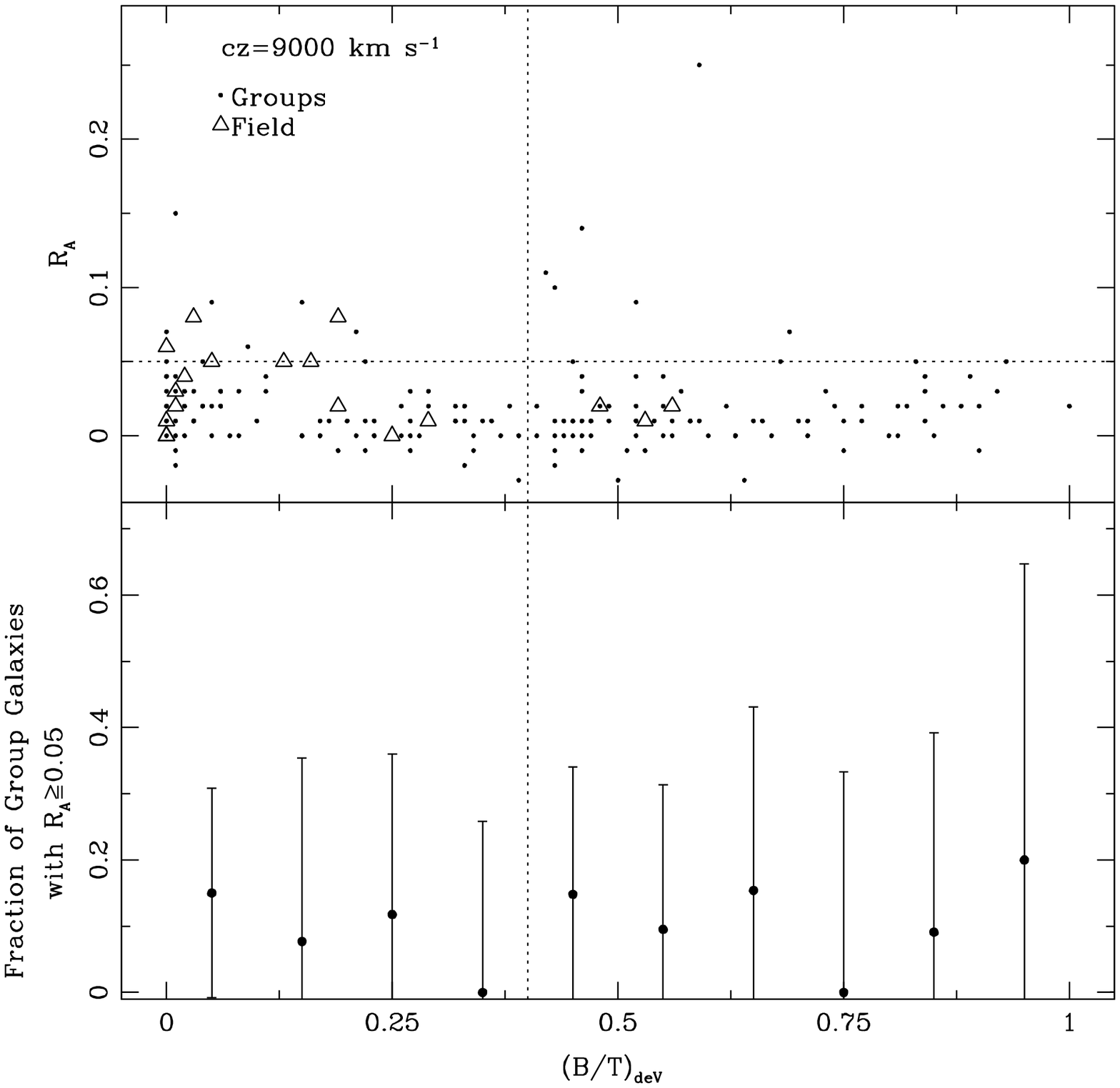}{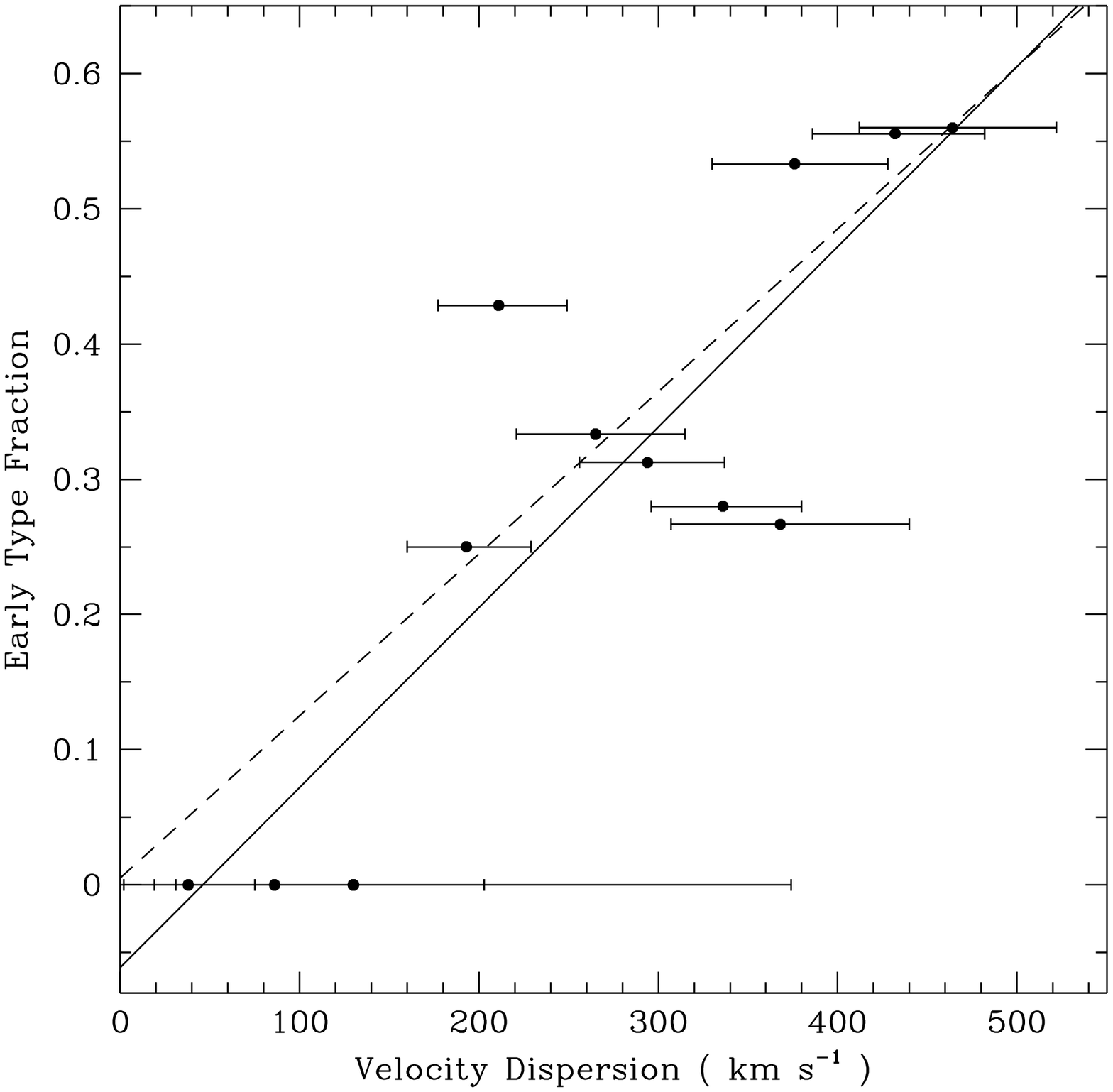}
\caption{\small {\it Left:}
Distribution of the asymmetry parameter $R_A$ with respect to the fraction of galaxy
light within the bulge component $B/T$ (Tran {\it et al.} 2001).  Galaxies with $R_A > 0.05$ are considered significantly
asymmetric.  Galaxies with $B/T > 0.4$ are classified bulge-dominated.
{\it Right:}  
Early type fraction vs. velocity dispersion for poor groups in Zabludoff \& Mulchaey 1998.
The dashed and solid lines are weighted and unweighted fits, respectively.
} 
\label{fig3}
\end{figure}

\section{Why Study Poor Groups?}

Poor groups are important to studies of cosmology and galaxy evolution
for reasons that include the ubiquity of the group environment, the
fact that clusters of galaxies evolve hierarchically from the
accretion of groups, and the likelihood of tidal interactions and
mergers among group members.  First, let
us consider the prevalence of groups.  More than half of the baryons
in the nearby universe lie in groups (Fukugita {\it et al.} 1998), the
intergroup medium may contribute to the population of high redshift
absorption line systems (Mulchaey {\it et al.} 1996), and group environment
plays a critical, if poorly-defined role, in the degree to which
gravitational lens galaxies distort the light of background objects.
In this last case, it is possible to estimate the fraction of lensing
galaxies that lie in groups (Keeton, Christlein, \& Zabludoff 2000).
The best lenses are the most massive galaxies, which tend to be early
types and to lie in the densest regions.  However, the increase in the
dwarf-to-giant galaxy ratio with environment density (\S3)
compensates for the morphology-density relation, causing the
distribution of environments of lens and non-lens galaxies to be
similar.  We estimate that at least $25\%$ of strong lensing galaxies
are group or rich cluster members.  As a consequence, the extended
dark halo of groups and the individual halos of other group galaxies
will significantly affect the lensing models and the
constraints they place on halo shapes and cosmological parameters like
$H_0$ and $\Lambda$.

Second, the accretion of groups by rich clusters provides an
opportunity to test the effects of cluster environment on galaxy
evolution by comparing groups in the field to groups (subclusters)
that have recently fallen into clusters.  By comparing the galaxy
morphologies and recent star formation histories for group galaxies in
and out of clusters, it is possible to quantify the degree to which
mechanisms such as ram pressure stripping or galaxy harassment
--- which are not as efficient in group environments -- affect the
evolution of cluster galaxies.  Preliminary evidence suggests that the
fraction and star formation histories of early types do {\it not}
differ significantly from field groups to subclusters, suggesting that
any external forces that drive galaxy evolution are not
cluster-dependent but group-dependent (Zabludoff \& Mulchaey 1998).
One such mechanism is galaxy-galaxy tidal encounters and mergers, which
are more efficient in poor groups than in clusters because the velocity
dispersion of a group is similar to that of an individual galaxy.

  

Third, as likely sites for interaction-driven galaxy evolution,
groups, especially the most dynamically young systems, are good
places to look for galaxies evolving at $z \sim 0$.  While
evidence suggests that the cores of rich clusters have old, mostly
unevolving galaxy populations, the beautiful HI
map of the M81 group on the conference T-shirt (Yun {\it et al.} 1994)
clearly shows that transforming interactions among galaxies can occur
in poor groups today.  Even though galaxy morphologies are frequently
less revealing at optical wavelengths, optical imaging is abundant and
worth examining for disturbed galaxies.  In Tran {\it et al.} (2001), we obtain
quantitative optical morphologies for galaxies in evolved, X-ray
luminous groups by fitting the bulge and disk, subtracting the best
model, and determining the asymmetric component of the residual light.
Figure 3 shows that $\sim 15\%$ of these galaxies, including both early and
late types, are significantly asymmetric.  The presence of obvious
tidal features in some of these galaxies suggests that
morphologically-disruptive interactions can take place even in
dynamically older groups.

While such interactions may be more common in lower velocity
dispersion, less evolved groups like the Local Group and the M81
system, there is indirect evidence that interactions have played a
role in most groups at some time.  For example,
Figure 3 shows a strong correlation between the early type fraction
and velocity dispersion of poor groups.
Either both quantities increase as groups evolve or the initial size
of the group potential determines the early type fraction for all
time.  Some support for the former scenario comes from 
the saturation point in early type fraction beyond which the morphological
compositions of groups and clusters are similar.  This saturation
point is about 500 km s$^{-1}$, the velocity dispersion that a group
must have such that an L$^*$ galaxy would have merged with another
member within a Hubble time.

  

\section{The Group Galaxy Luminosity Function}

\begin{figure}
\plottwo{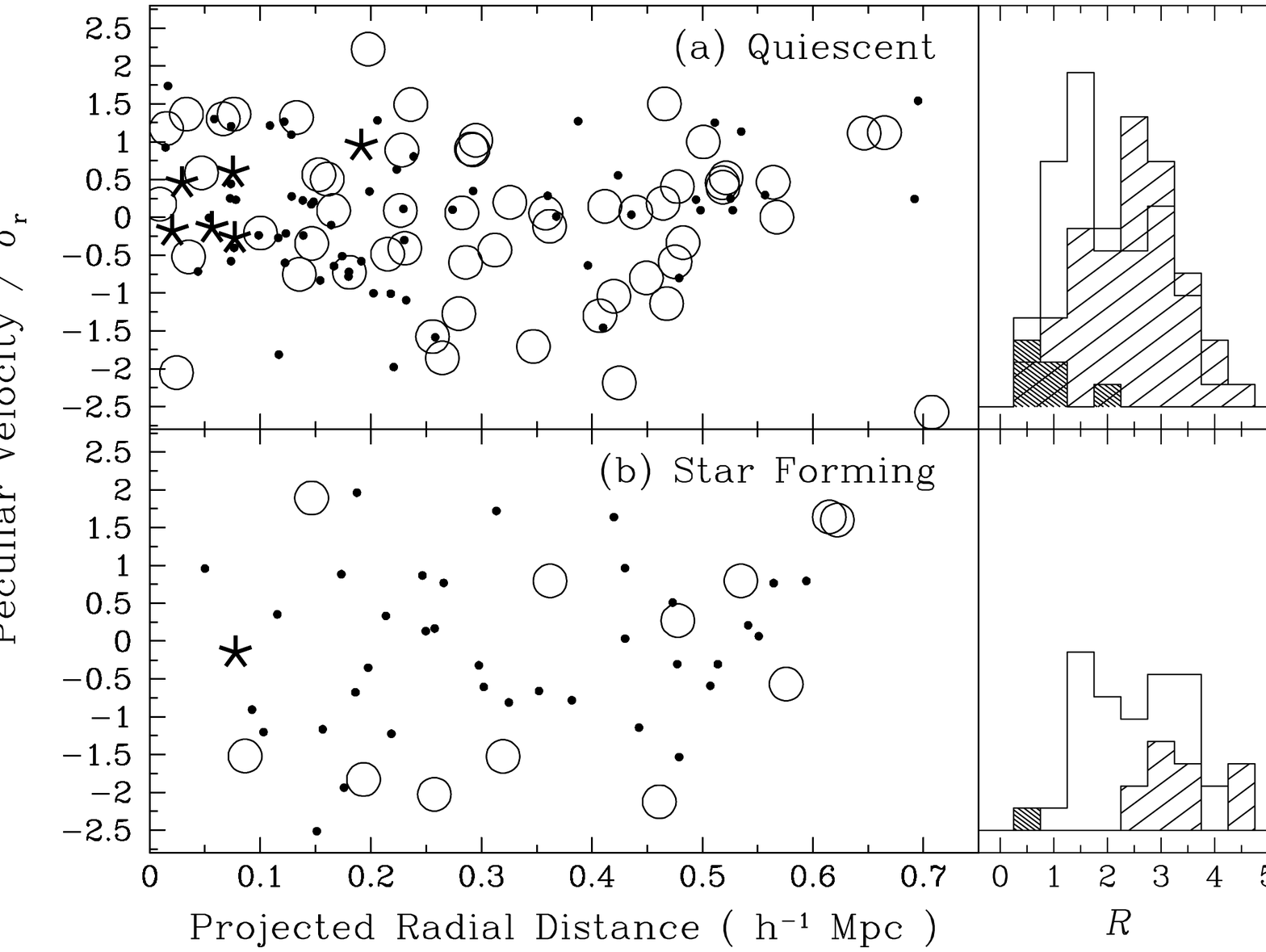}{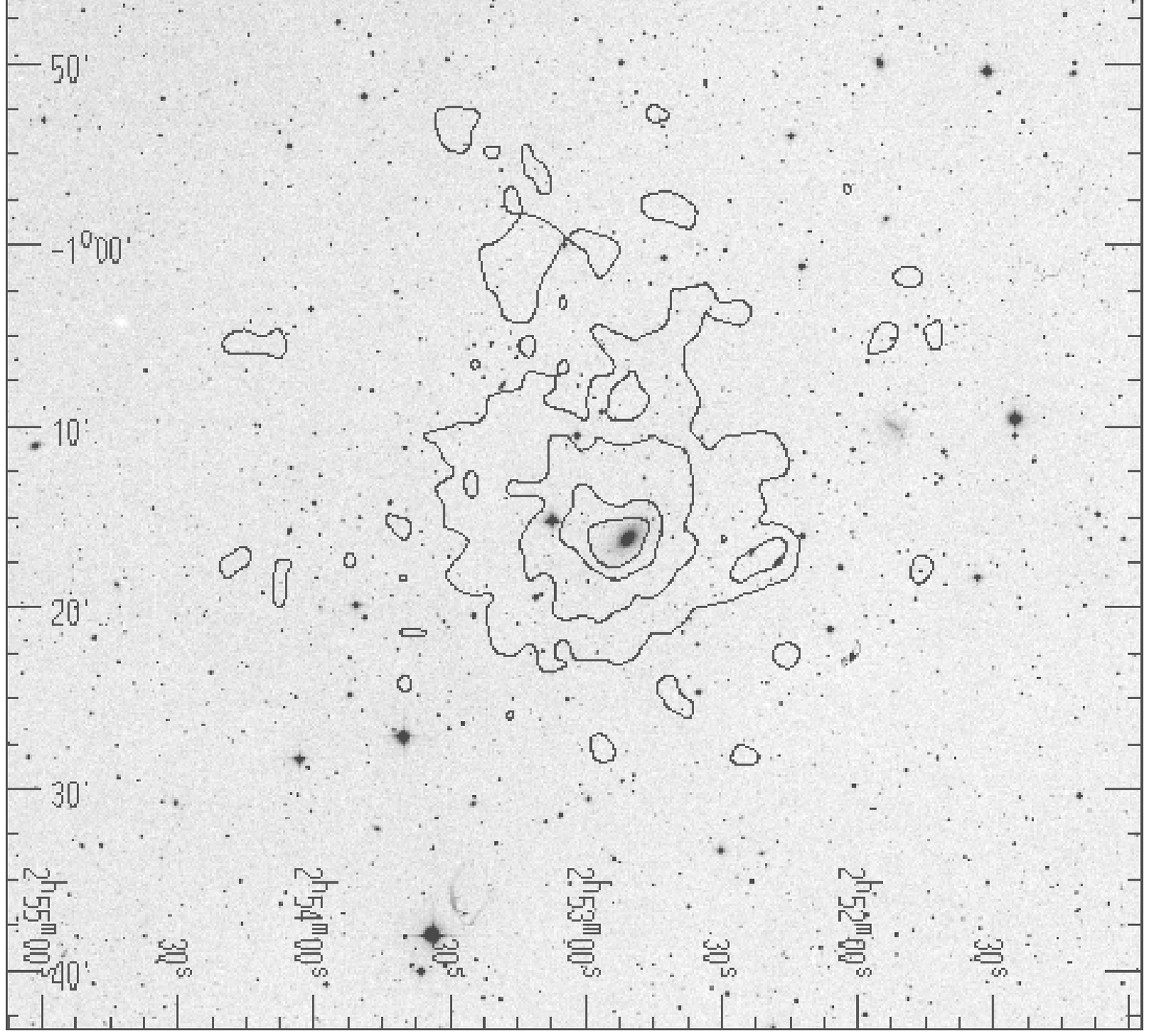}
\vskip -2. cm
\caption{\small {\it Left:}
Velocity offset vs. projected radial offset 
from the group centroid of (a) 123 quiescent and (b) 49 star-forming
group members from six poor groups.
The asterisks are the brightest group members
($M_R \leq M_R^*$). The open circles are the
giants defined by $M_R^* < M_R \leq  M_R -19 + 5$ log $h$. The filled circles are the
dwarfs ($-19 + 5$ log $h$ $< M_R \leq -17 + 5$ log $h$).
Also plotted is the distribution of $R$, 
the quadrature sum of the x- and y-axis offsets of each galaxy,
for the brightest (heavily shaded), giant (shaded), and dwarf populations
(unshaded). The different R distributions suggest that the three populations
occupy distinct orbits.
{\it Right:}  
A Digital Sky Survey image
of the isolated elliptical NGC 1132 (greyscale) with the ASCA X-ray emission (contours) superimposed
(Mulchaey \& Zabludoff 1999).}
\label{fig4}
\end{figure}

If interactions can alter the morphology of group galaxies, they can
also affect galaxy luminosity.  We might expect the ratio of
dwarf-to-giant galaxies ($D/G$) to be dependent on both the initial
luminosity distribution when most galaxies formed and the subsequent
evolution of galaxy luminosity and number density.  In both cases, the
environment may play a role.  For example, standard biased galaxy
formation models (White {\it et al.} 1987) predict that giant
galaxies are more likely than dwarfs to form in dense regions.
Subsequent evolution via galaxy-galaxy mergers, tidal interactions,
ram pressure stripping, or other factors would alter $D/G$ in a direction
and to a degree that are dependent on the dominant mechanism.  To
test the dependence of $D/G$ on environment, we investigate whether
$D/G$ varies among poor groups with different galaxy number densities,
$within$ groups with radius, and from the field to groups to rich clusters.

We find that the galaxy luminosity function is not universal --- the
group with the faintest X-ray emission and lowest galaxy number
density has a significantly lower $D/G$ than the other six systems in
our sample.  Furthermore, the average $D/G$ for the five X-ray
luminous groups is larger than that derived from the Las Campanas
Redshift Survey field (Lin {\it et al.} 1996) and consistent with
that of rich clusters (cf. Trentham 1997)\footnote{The
similarity of $D/G$ for groups and clusters
once groups reach a large enough gravitational well depth
is reminiscent of the saturation of the early type fraction-velocity
dispersion relation noted in the previous section.}.
Within the
composite group, $D/G$ drops significantly with increasing radius (and
thus decreasing density).  These three trends are in the same sense
and suggest that, at least in some environments, dwarfs are more
biased than giants with respect to dark matter.  Thus more than just
standard biased galaxy formation is at work (Zabludoff \&
Mulchaey 2000).

Evidence that the group dwarf and giant populations formed at different times
is presented in Figure 4.  The phase-space diagram
shows that the dwarfs and giants have not mixed, {\it i.e.}, lie on
different orbits.  
Possible explanations for this result and the
trends in $D/G$ summarized above include inefficient galaxy formation
({\it e.g.}, giants form less efficiently in denser environments),
increases in the satellite-to-primary ratio through the mergers of
giant galaxies, and dwarf formation in the tidal tails of giant merger
remnants (Barnes \& Hernquist 1992).

Additional support that $D/G$ evolves comes from X-ray and
optical observations of the isolated elliptical NGC 1132 (Figure 4).
Numerical simulations predict that some poor groups of galaxies have
merged by the present epoch into giant ellipticals (Barnes 1989).
The extent ($\sim 250$ kpc $h^{-1}$), temperature ($\sim 1$ keV), metallicity ($\sim 0.25$ solar),
and luminosity ($\sim 2.5 \times 10^{42}$ $h^{-2}$ ergs s$^{-1}$) of NGC 1132's X-ray halo are
comparable with those of poor group halos.  The total mass inferred
from the X-ray emission, $\sim 1.9 \times 10^{13}$ $h^{-1}$ M$_\odot$, is also like that of an X-ray
detected group.  Optical imaging uncovers a dwarf galaxy population
clustered about NGC 1132 that is consistent in number density and
projected radial distribution with that of an X-ray group.  The
similarities of NGC 1132 to poor groups in both the X-ray band and at
the faint end of the galaxy luminosity function, combined with the
deficit of luminous galaxies in the NGC 1132 field, are compatible
with the merged group picture. Another possibility is that the NGC
1132 system is a ``failed'' group ({\it i.e.}, a local overdensity in which
other bright galaxies never formed).

\begin{figure}
\plottwo{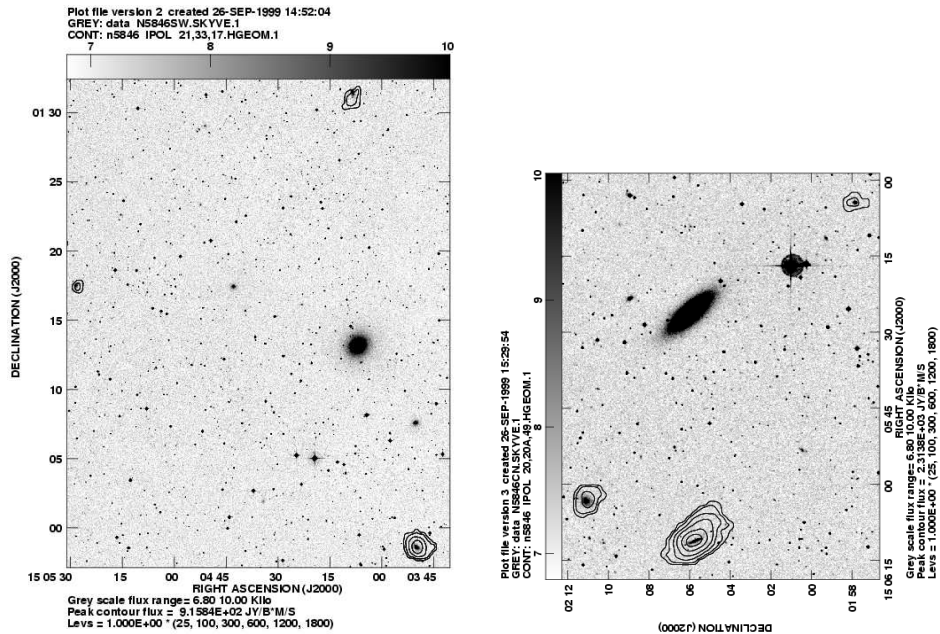}{20,20A,49}
\caption{\small 
Two images from our VLA mosaic survey of the NGC 5846 poor group showing several HI detections (contours) 
of group members.
}
\label{fig5}
\end{figure}

\section{Cold Gas in Groups}

In the previous sections, we discussed the galaxies and hot gas in
poor groups, but what about the cold gas?  HI studies of groups can
tell us both about interactions among galaxies --- witness the M81
group --- and about galaxy formation efficiency in groups.  Our
on-going HI survey with the VLA (Zabludoff with van Gorkom, Wilcots,
\& Mulchaey) targets not just group members, but the entire $1.5
\times 1.5$ square degree field of each group with 36 pointed mosaic
tiles.  Of the two groups examined so far,
we have made $\sim 15$ HI detections of galaxies with a
range of optical types and luminosities (Figure 5).

Our preliminary results include an increase in the HI extent of
galaxies with projected radius from the group center.  However, this
trend might be due to the increasing spiral fraction with projected
radius --- we are currently comparing the magnitudes of these two
effects.  Most of our detections are near the $\sim 10^7$ M$_\odot$ limit
of the survey, so many other detections may be possible at fainter
limits.  We observe no obvious galaxy-galaxy interactions like those
in M81, but fewer HI-rich interactions are expected in these
more dynamically evolved,
X-ray detected, high early type fraction groups.
All HI clouds lie at the positions of
galaxies, {\it i.e.}, there are no ``free'' clouds down to our
detection limit that could be associated with evolution of group as
whole (cf. Blitz {\it et al.} 1999) instead of individual galaxies.

\section{Conclusions}

Groups are where a lot of the action is (or was) in
galaxy evolution.  There are optical and HI signatures of on-going
galaxy-galaxy interactions, not only in M81,
but also to a lesser degree in more dynamically evolved, X-ray
luminous groups.  There is also indirect
evidence that groups, and the mergers and tidal encounters 
likely to occur in such low velocity dispersion systems, played an
important role in the past evolution of galaxies.  The galaxy
populations of X-ray luminous poor groups and rich clusters, including
the early type fraction, fraction of early types forming stars, and
dwarf-to-giant ratio, are surprisingly similar.  These results suggest
that group-dependent, not cluster-dependent, mechanisms like
galaxy-galaxy interactions drive the evolution of those galaxies that
begin as members of groups accreted by clusters\footnote{Even the
$\sim 400$ km s$^{-1}$ upper limit on the velocity dispersion of a
giant elliptical galaxy is consistent with the velocity dispersion of
a group --- is this limit due to the initial formation of such
galaxies in group, not cluster, environments?  Further support for
this idea comes from the tendency of the brightest giant ellipticals
to lie in the spatial and kinematic centers of groups and subclusters,
not clusters (Beers \& Geller 1983; Zabludoff \& Mulchaey 1998).}.

Additional evidence for environment-dependent evolution is the
increase in the dwarf-to-giant ratio with density, at least up to the
densities of X-ray luminous groups and clusters, indicating that more
than standard biased galaxy formation is at work.  The dwarf and giant
galaxies in groups have not mixed, suggesting that one population
formed later.  This evolution in $D/G$ might be due to galaxy-galaxy
interactions; a compelling example is NGC 1132, an isolated giant
elliptical with all the hallmarks of a merged group.  On-going HI
studies with the VLA have the potential to place critical constraints
on the interaction history and galaxy formation efficiency in such
groups.





\begin{references}
{\small
\reference Barnes, J.E. \& Hernquist, L. 1992, \araa, 30, 705
\reference Barnes, J.E. 1989, Nature, 338, 123
\reference Beers, T. \& Geller, M. 1983, \apj, 274, 491
\reference Blitz, L., Spergel, D., Teuben, P., Hartmann, D., \& Burton, W.B. 1999, \apj, 514, 818
\reference Fukugita, M., Hogan, C. J., Peebles, P. J. E. 1998, \apj, 503, 518
\reference Keeton, C.R., Christlein, D., \& Zabludoff, A.I. 2000, \apj, 545, 129
\reference Lin, H., Kirshner, R.P., Shectman, S.A., Landy, S.D., Oemler, A., Tucker, D. L., Schechter, P. L. 1996, \apj, 464, 60
\reference Mulchaey, J.S. \& Zabludoff, A.I. 1999, \apj, 514, 133
\reference Mulchaey, J.S., Mushotzky, R.F., Burstein, D., Davis, D.S. 1996, \apj, 456, 5
\reference Tran, K.-V., Simard, L., Zabludoff, A.I., \& Mulchaey, J.S. 2001, \apj, in press 
\reference Trentham, N. 1997, \mnras, 290, 334
\reference White, S., Davis, M., Efstathiou, G., \& Frenk, C. 1987, Nature, 330, 451
\reference Yun, M. S., Ho, P. T. P., Lo, K. Y. 1994, Nature, 372, 530
\reference Zabludoff, A.I. \& Mulchaey, J.S. 2000, \apj, 539, 136
\reference Zabludoff, A.I. \& Mulchaey, J.S. 1998, \apj, 496, 39
}
\end{references}
\end{document}